\documentclass[a4paper,reprint,12pt,aps,tightenlines,onecolumn,notitlepage,superscriptaddress]{revtex4-1}
\usepackage{amsmath}
\usepackage{amssymb}
\usepackage{cancel}
\usepackage{stmaryrd}
\usepackage{amsfonts}
\usepackage[T1]{fontenc}
\usepackage{bm}
\usepackage{color}
\usepackage{graphicx}
\DeclareSymbolFont{operators}{OT1}{cmr}{m}{n}
\DeclareSymbolFont{letters}{OML}{cmm}{m}{it}
\DeclareSymbolFont{symbols}{OMS}{cmsy}{m}{n}
\DeclareSymbolFont{largesymbols}{OMX}{cmex}{m}{n}
\usepackage[hyperindex,breaklinks]{hyperref}
\hypersetup{
    pdfborder={0 0 0},
    allbordercolors={0 0 0},
    bookmarks=true,         
    unicode=false,          
    pdftoolbar=true,        
    pdfmenubar=true,        
    pdffitwindow=false,     
    pdfstartview={FitH},    
    pdfcreator={Eero Hirvijoki},
    pdfproducer={pdflatex},
    pdfnewwindow=true,
    colorlinks=true,
    linkcolor=blue,
    citecolor=blue,
    filecolor=blue,
    urlcolor=blue,  
}
\usepackage{array}

\newcommand{\fa}{\mathcal{A}}
\newcommand{\fb}{\mathcal{B}}
\newcommand{\fc}{\mathcal{C}}
\newcommand{\fe}{\mathcal{E}}
\newcommand{\ff}{\mathcal{F}}
\newcommand{\fg}{\mathcal{G}}
\newcommand{\fh}{\mathcal{H}}
\newcommand{\fm}{\mathcal{M}}
\newcommand{\fs}{\mathcal{S}}
\newcommand{\fp}{\mathcal{P}}
\newcommand{\fu}{\mathcal{U}}

\newcommand{\matrixform}[1]{\ensuremath{\mathbb{#1}}}

\newcommand{\MG}{\matrixform{G}}

\newcommand{\ML}{\matrixform{L}}

\newcommand{\MV}{\matrixform{V}}

\newcommand{\bracket}[1]{\ensuremath{\left < #1 \right >}}

\DeclareFontFamily{OT1}{pzc}{}
\DeclareFontShape{OT1}{pzc}{m}{it}{<-> s * [1.10] pzcmi7t}{}
\DeclareMathAlphabet{\mathpzc}{OT1}{pzc}{m}{it}

\makeatother

\begin{document}

\title{Metriplectic particle-in-cell integrators for the Landau collision operator}

\author{Eero Hirvijoki}
\affiliation{Princeton Plasma Physics Laboratory, Princeton, New Jersey 08543, USA}
\email{ehirvijo@pppl.gov; eero.hirvijoki@gmail.com}

\author{Michael Kraus}
\affiliation{Max-Planck-Institut f\"ur Plasmaphysik, Boltzmannstra\ss e 2, 85748 Garching, Deutschland}
\email{michael.kraus@ipp.mpg.de}
\affiliation{Technische Universit\"at M\"unchen, Zentrum f\"ur Mathematik, Boltzmannstra\ss e 3, 85748 Garching, Deutschland}
\email{michael.kraus@ipp.mpg.de}
\author{Joshua W. Burby}
\affiliation{Courant Institute of Mathematical Sciences, New York, New York 10012, USA}
\email{joshua.burby@cims.nyu.edu}

\date{\today}

\begin{abstract}
In this paper, we present a new framework for addressing the nonlinear Landau collision operator in terms of particle-in-cell methods. We employ the underlying metriplectic structure of the collision operator and, using a macro particle discretization for the distribution function, we transform the infinite-dimensional system into a finite-dimensional time-continuous metriplectic system for advancing the macro particle weights. Temporal discretization is accomplished using the concept of discrete gradients. The conservation of density, momentum, and energy, as well as the positive semi-definite production of entropy in both the time-continuous and the fully discrete system is demonstrated algebraically. The new algorithm is fully compatible with the existing particle-in-cell Poisson integrators for the Vlasov-Maxwell system.
\end{abstract}


\maketitle

\section{Introduction}
For several decades, the particle-in-cell framework \cite{Dawson:1962} has delivered algorithms to simulate the Vlasov-Maxwell system and its simplifications. Recently, the framework has advanced greatly \cite{Chen:2011jf,Markidis:2011ep,Chacon:2013cz,Chen:2014eh}, climaxing in the so-called variational \cite{Squire:2012} and Poisson integrators \cite{Qin:2016,He:2016,Kraus:2016:GEMPIC,Burby-2017:finite-dimensional} that preserve the mathematical structure of the Vlasov-Maxwell equations also on the discrete level. For a thorough review on the subject, see Ref.~\cite{Morrison:2017} and references therein. Despite this tremendous success, the proper treatment of small-angle Coulomb-scattering effects within the particle-in-cell framework still is an open problem. The binary collision algorithm, pioneered by Takizuka and Abe \cite{Takizuka-Abe:1977} and mastered by Nanbu~\cite{Nanbu:1997,Nanbu-Yonemura:1998,Bobylev-Nanbu:2000}, preserves conservation laws only if each macro particle has an equal, fixed weight throughout the simulation. The other existing Monte Carlo approach, that exploits the connection of the Landau collision operator and stochastic differential equations, allows arbitrary macro-particle weights but requires ad-hoc enforcement of the conservation laws~\cite{Dimits-Cohen:1994,Wang-deltaf:1999}. These issues in the Monte Carlo collision algorithms void the mathematical rigor invested into  treating the Vlasov-Maxwell part. 

Fortunately, the problem of collisions can also be approached from a different perspective. While Hamiltonian flow of infinitesimal phase-space volume elements is conveniently interpreted as pushing macro particles according to their Hamiltonian trajectories, collisions can be understood to either increase or decrease the phase-space density in the infinitesimal volume element, without disturbing the Hamiltonian trajectories. This interpretation leads to a collision algorithm that is typically implemented in three steps~\cite{Yoon:2014:POP}. The macro particles are first projected onto a mesh to obtain either a finite-element or a finite-volume representation of the distribution function. The distribution function is then evolved according to the Landau collision operator on the mesh, an operation for which conservative schemes exist \cite{Yoon:2014:POP,Taitano:2015JCP,Hager:2016:JCP,Hirvijoki:2017ei}. Finally, new particle weights are interpolated from the distribution function into existing macro particle locations. While the approach is physically intuitive, its mathematical origin is unclear as no time-continuous equation for the change in the particle weights has been derived. Depending on the discretization on the mesh, the transfer of weights from the mesh representation to the macro particles may also suffer from interpolation error, typically leading to loss of energy conservation \cite{Yoon:2014:POP}. Finally, no discrete H-theorem has been shown to exist.

In this paper, we take a structured approach to changing the macro particle weights via collisions and build a mathematical foundation for the existing three-step collision algorithms. We do this by proposing a new class of particle-in-cell integrators for the Landau operator which employ the perhaps less familiar, so-called metriplectic formulation of the collision operator \cite{Morrison:1986vw,Kraus-Hirvijoki-2017}. Using a macro particle discretization for the distribution function, we transform the infinite-dimensional metriplectic formulation of the Landau collision operator into a finite-dimensional time-continuous metriplectic system for advancing the macro particle weights. Temporal discretization of the weight equations is then accomplished using the concept of discrete gradients \cite{Quispel:1996,McLachlan:1999,QuispelMcLaren:2008,MansfieldQuispel:2009,CohenHairer:2011}. The conservation of density, momentum, and energy, as well as the positive semi-definite production of entropy in both the time-continuous and the fully discrete system are demonstrated algebraically. Finally, as our finite-dimensional expression for entropy remains a Casimir of the finite-dimensional Poisson brackets discussed in Refs.\cite{Qin:2016,He:2016,Kraus:2016:GEMPIC,Burby-2017:finite-dimensional}, the presented metriplectic integrator for the Landau collision operator is fully compatible with the existing particle-in-cell Poisson integrators for the Vlasov-Maxwell system.

The new results we present have roots in our previous works. In Ref.~\cite{Hirvijoki:2017ei} we realized the potential of direct finite-element discretizations of the Landau collision operator and, in Ref.~\cite{Kraus-Hirvijoki-2017}, this potential was exploited further to construct finite-element discretizations of the metriplectic formulation. Based on the accumulated understanding, we can finally approach the collision operator also from the particle-in-cell perspective. To translate our reasoning efficiently, the paper focuses on explaining our choices with respect to the discretization, assuming the generic properties and the basic components of the metriplectic Vlasov-Maxwell-Landau system to be consulted from previous publications \cite{Morrison:1986vw,Kraus-Hirvijoki-2017}. Although the presented results open up a new and exciting research direction, they provide only the first step. For example, one of the remaining open questions is to find a metriplectic algorithm that can be algebraically shown to retain the strict non-negativity of the macro particle weights. We leave this and other such challenges to future work.
\section{Metriplectic formulation of collisions}
Previously, it has been shown that the evolution of arbitrary functionals $\fu[f,E,B]$, corresponding to dynamics of the Vlasov-Maxwell-Landau system with $f$ a phase-space particle distribution function and $(E,B)$ the electromagnetic fields, can be obtained from the so-called metriplectic formalism. In terms of the Vlasov-Maxwell Poisson bracket $\{\cdot,\cdot\}$ \cite{Morrison:1980,Weinstein:1981,Marsden:1982} and a negative semi-definite metric bracket $(\cdot,\cdot)$ \cite{Morrison:1986vw,Kraus-Hirvijoki-2017} representative of the collision operator, the time evolution in the system is given by 
\begin{align}
\frac{d\fu}{dt}=\{\fu,\ff\}+(\fu,\ff),
\end{align}
with the free energy functional $\ff[f,E,B]=\fh[f,E,B]-\fs[f]$ determined in terms of the Vlasov-Maxwell Hamiltonian $\fh$ and an entropy $\fs$. If the metric bracket and the entropy functional are dropped, structure-preserving particle-in-cell algorithms exist to address the dissipation-free evolution of the Vlasov-Maxwell system, as discussed in the introduction. In this paper, we thus focus on the dissipative, collisional contributions that manifest in the metric bracket $(\cdot,\cdot)$. For purposes of keeping the notation as simple as possible, the analysis represented here shall focus on the single-species Landau operator. The generalization to multiple species is straightforward, though requires more index-keeping. We shall also assume normalized units for convenience.

The collisional evolution in the Vlasov-Maxwell-Landau system is provided by the nonlinear Landau operator \cite{Landau:1936}
\begin{align}\label{eq:landau-operator}
C(f)=\frac{\partial}{\partial v}\cdot\int Q(v-v')\cdot \left(f(x,v')\frac{\partial f(x,v)}{\partial v}-f(x,v)\frac{\partial f(x,v')}{\partial v'}\right) dv'.
\end{align}
The dyad $Q(\xi)=|\xi|^{-1}(1-\xi\xi/|\xi|^2)$ serves as a scaled projection matrix, and the collision operator acts locally in configuration space $x$. We shall use $z=(x,v)$ to denote phase-space coordinates. The generic form of the negative semi-definite bracket, capable of reproducing the collisional dynamics, is 
\begin{align}
(\fa,\fb)&=\int\int \Gamma(\fa;z,z')\cdot W(f;z,z')\cdot \Gamma(\fb;z,z')dz dz',
\end{align}
where the vector $\Gamma$ is given by
\begin{align}
\Gamma(\fa;z,z')=\left(\frac{\partial}{\partial v}\frac{\delta \fa}{\delta f(z)}-\frac{\partial}{\partial v'}\frac{\delta \fa}{\delta f(z')}\right),
\end{align}
and the dyad $W$ is given by
\begin{align}
&W(f;z,z')=-\frac{1}{2}M(f)(z)M(f)(z')Q(v-v')\delta(x-x'),
\end{align}
with $\delta(x-x')$ ensuring the locality of the collisions in the $x$-space. The metric bracket corresponding to the Landau operator and Maxwell-Boltzmann statistics is recovered with the choices $M(f)=f$ and $\fs[f]=\int s(f)dz$ with $s(f)=-f\ln f$. Fermi-Dirac statistics and the corresponding small-angle scattering operator would be obtained with different choices for $s(f)$ and $M(f)$~\cite{Morrison:1986vw}. Effectively, $s$ and the eigenvector of $Q$ corresponding to zero eigenvalue together determine the particle statistics and the equilibrium state while the map $M$ can be used to tune the collision operator. 

Given the metric bracket, the entropy functional corresponding to the Maxwell-Boltzmann statistics, and the Vlasov-Maxwell Hamiltonian
\begin{align}
\fh=\frac{1}{2}\int |v|^2f dz+\frac{1}{2}\int( |E|^2+|B|^2)dx,
\end{align}
it is straightforward to obtain the collisional dynamics for the distribution function: choosing a functional $\fg[f]=\int g(z)f(z)dz$, with respect to an arbitrary test function $g$, provides exactly the weak formulation of \eqref{eq:landau-operator}. Verification of this is left as an exercise for an enthusiastic reader. The conservation properties of the Landau collision operator are reproduced in the form of so-called Casimir invariants $\fc_i$ for which $(\fc_i,\fa)=0$ for any given $\fa$. The metric bracket has three classes of Casimirs corresponding to mass density, kinetic momentum density, and kinetic energy density,
\begin{align}
\fc_{\fm}=\int g_{\mathcal{M}}(x) f(z) dz, &&
\fc_{\fp}=\int g_{\mathcal{P}}(x)v f(z) dz, &&
\fc_{\fe}=\frac{1}{2}\int g_{\mathcal{E}}(x)|v|^2 f(z) dz,
\end{align}
with $g_{\mathcal{M}}(x)$, $g_{\mathcal{P}}(x)$, and $g_{\mathcal{E}}(x)$ arbitrary functions of $x$-space. The first two follow from $\Gamma(\fc_{\fm})=0$ and $\delta(x-x')\Gamma(\fc_{\fp})=0$, and the last from $\delta(x-x')\Gamma(\fc_{\fe})=g_{\mathcal{E}}(x)(v-v')$ and the property $\xi\cdot Q(\xi)=0$. 
Furthermore, as the metric bracket is negative semi-definite, the free energy is dissipated and entropy produced,
\begin{align}
\frac{d\ff[f,E,B]}{dt}=(\ff,\ff)\leq 0, && 
\frac{d\fs[f]}{dt}=-(\fs,\fs)\geq 0.
\end{align}
The latter follows from the conditions that the Vlasov-Maxwell Hamiltonian is a Casimir of the metric bracket and the entropy is a Casimir of the Vlasov-Maxwell Poisson bracket. 

Our target in this paper is (a) to introduce a macro particle representation for the distribution function compatible with existing structure-preserving particle-in-cell algorithms for the Vlasov-Maxwell part, and (b) to find a finite-dimensional approximation for the metric bracket for advancing the macro particle weights. While different discretizations of the metric bracket are possible in principle, we seek a discretization that mimics the Casimir structure and negative semi-definiteness of the continuous metric bracket. In particular, the discrete metric bracket presented in~\cite{Kraus-Hirvijoki-2017} is \emph{not} sufficient to address (b) by itself. The latter discrete bracket applies to a special class of finite-element discretizations of the distribution function, while here we focus on macro particle discretizations with time-varying weights.


\section{Spatial discretization with macro particles}
In classical particle-in-cell methods, the distribution function is typically represented in the form 
\begin{align}\label{eq:pic_f_ideal}
f_h(x,v,t)=\sum_p w_p \Psi(x-x_p(t))\Phi(v-v_p(t)),
\end{align}
with $\Psi$ and $\Phi$ positive, fixed shape functions. The shape functions are normalized according to $\int \Psi(x-x_p)dx=1$ and $\int \Phi(v-v_p) dv=1$, meaning that the weights $w_p$ effectively correspond to the number of real particles the macro particles represent, i.e. the number of actual particles in the phase space region defined by the support of the (product of the) shape functions. In the Vlasov-Maxwell system, the resulting equations in a particle-in-cell algorithm advance the degrees of freedom $z_p=(x_p,v_p)$ while not changing the weights $w_p$. As discussed in the introduction, we shall address the collisional evolution of the system assuming that the metric bracket operates only on the macro particle weights, leaving the phase-space locations $z_p$ intact. Thus, in the metric bracket, the discrete representation for the distribution function is interpreted according to
\begin{align}\label{eq:pic_f}
f_h(x,v,t)=\sum_p w_p(t) \Psi(x-x_p)\Phi(v-v_p),
\end{align}
instead, with $w=\{w_p\}_{p}$ now being the degrees of freedom in the system. 
These two different interpretations, \eqref{eq:pic_f_ideal} and \eqref{eq:pic_f}, facilitate efficient splitting schemes, where the Hamiltonian and the collisional part of the Vlasov-Maxwell-Landau system are solved in an alternating fashion.

A key insight for the discretization of the metriplectic bracket is that a functional $\fa[f]$, when evaluated with respect to the discrete distribution function $f_h$, becomes a function $\fa_h(w)$ of the degrees of freedom $w$, defined via the scalar invariance relation 
\begin{align}\label{eq:scalar-invariance}
\fa[f_h]=\fa_h(w).
\end{align}
A finite-dimensional bracket, acting on functions $\fa_h$ and $\fb_h$ of the degrees-of-freedom $w$, could thus be obtained by evaluating the infinite-dimensional bracket with respect to the discrete distribution function,
\begin{align}
(\fa_h,\fb_h)_h(w)\equiv(\fa,\fb)[f_h],
\end{align}
where $\fa,\fb$ are functionals on the infinite-dimensional $f$-space that agree with $\fa_h,\fb_h$ when evaluated on distribution functions of the form given in \eqref{eq:pic_f}. In other words, $\fa,\fb$ should be chosen according to some well-defined prescription that ensures $\fa(f_h)=\fa_h(w)$. In this way the collisional dynamics of the finite-dimensional system would be provided by the equation
\begin{align}\label{eq:finite-dynamics}
\frac{d\fu_h}{dt}=(\fu_h,\ff_h)_h.
\end{align}

While this appears to be a simple recipe, several problems need to be solved.
First, 
in the current context, it is not immediately clear how to relate functional differentiation with respect to $f$ to functional differentiation with respect to $w$.
Second, requiring the finite-dimensional bracket to satisfy discrete versions of the fundamental conservation laws (mass, momentum, energy) complicates the process.
Finally, the finite-dimensional expression for entropy functional has to remain a Casimir of the existing finite-dimensional Poisson brackets of the Vlasov-Maxwell part.
A possible solution to the first problem is proposed below
and several other solutions are anticipated to exist.
Finding a solution to the second problem is guided by careful investigation of the origin of the conservation laws in the infinite-dimensional case, i.e., the conditions $\Gamma(\fc_\fm)=0$, $\delta(x-x')\Gamma(\fc_\fp)=0$, and $\delta(x-x')\Gamma(\fc_\fe)=g_{\mathcal{E}}(x)(v-v')$. If these conditions can be reproduced for the finite-dimensional representation of the distribution function, as demonstrated in Section~\ref{sec:finite-bracket}, the finite-dimensional bracket will have the desired conservation properties. 
The solution to the last issue, the choice of entropy, relies on approximating the entropy functional with a function that depends only on the macro particle weights. This approximation and justification for it will be discussed in detail in Section~\ref{sec:entropy}. 

In previous work~\cite{Hirvijoki:2017ei,Kraus-Hirvijoki-2017} on discretizing the Landau collision operator, we have observed that any finite-element or finite-volume basis used to represent $f$ should be chosen so that it can exactly reproduce the functions $(1,v,|v|^2)$.
The reasoning that lead to this conclusion follows from the scalar-invariance condition for functionals, and the fact that the functional derivatives of the infinite-dimensional mass, kinetic momentum, and kinetic energy density functionals result in $(1,v,|v|^2)$ multiplied by corresponding arbitrary functions $(g_{\mathcal{M}}(x),g_{\mathcal{P}}(x),g_{\mathcal{E}}(x))$. 
In what follows next, we exploit this observation in the context of a macro particle representation for $f$. 

In discretizing functional derivatives within the metric bracket, we choose to approximate generic functions $h(x,v)$ according to 
\begin{align}
h(x,v)=\sum_{i\in I}\sum_{j\in J^i}\hat{h}^i_j\mathbf{1}_{\Omega_i}(x)\mathsf{V}^i_j(v),
\end{align}
where $\Omega=\cup_{i\in I}\Omega_i$ denotes the computational domain in $x$-space, $\mathbf{1}_{\Omega_i}(x)$ is the standard indicator function being one if $x\in\Omega_i$ and zero otherwise, and $\{\Omega_i\}_{i\in I}$ represent the meshing of the $x$-space domain $\Omega$. 
The notation is chosen so that degrees-of-freedom are denoted with a "hat" and the superscript and subscript in the degrees-of-freedom denote the $x$- and $v$-space respectively. For each $x$-space domain $\Omega_i$, we choose a (possibly different) set of velocity-space basis functions $\{\mathsf{V}^i_{j}\}_{j\in J^i}$, which are required to satisfy the following conditions
\begin{align}\label{eq:basis-conditions}
1=\sum_{j\in J^i}\hat{1}^i_{j}\mathsf{V}^i_j(v), && v=\sum_{j\in J^i}\hat{v}^i_{j}\mathsf{V}^i_j(v), &&
\frac{1}{2}|v|^2=\sum_{j\in J^i}\hat{e}^i_{j}\mathsf{V}^i_j(v), && \forall\, i\in I,
\end{align}
for some $(\hat{1}^i_j,\hat{v}^i_j,\hat{e}^i_j)$. This is equivalent to requiring that the functions $(1,v,|v|^2)$ are contained in the span of $\{\mathsf{V}_j^i\}_{j\in J^i}$ for each $i\in I$. 
A simple example of a valid choice of $\mathsf{V}_j^i$ would be a discontinuous Galerkin basis of locally second-order polynomials in velocity, with possibly a different mesh for each domain $\Omega_i$. These definitions also allow us to conveniently approximate functions of only the variable $x$ according to 
\begin{align}
h(x)=\sum_{i\in I}\hat{h}^i\mathbf{1}_{\Omega_i}(x),
\end{align}
with the coefficients $\hat{h}^i$ corresponding to the average values of $h(x)$ in the domains $\Omega_i$. 

We now turn to defining discrete functional derivatives. In varying arbitrary functionals with respect to the discrete distribution function, the scalar invariance condition \eqref{eq:scalar-invariance} provides us with the following compatibility constraint on any definition we might choose:
\begin{align}\label{eq:fd_constraint}
\delta \fa[f_h]=\sum_p\int \frac{\delta \fa}{\delta f}\bigg|_{f_h}\delta w_p\Psi(x-x_p)\Phi(v-v_p)dz
\equiv\sum_p\frac{\partial \fa_h}{\partial w_p}\delta w_p=\delta \fa_h(w).
\end{align}
If we now demand that $\delta\fa/\delta f |_{f_h}$ is an element of our phase-space Galerkin basis, i.e.
\begin{align}
\frac{\delta \fa}{\delta f}\bigg|_{f_h}=\sum_{i\in I}\sum_{j\in J^i}\hat{a}^i_j\mathbf{1}_{\Omega_i}(x)\mathsf{V}^i_j(v),
\end{align}
then there is a unique way to express the $\hat{a}^i_j$ in terms of the $\partial\fa/\partial w_p$ such that \eqref{eq:fd_constraint} is satisfied.
We choose the $x$-space shape function for the macro particles to be
\begin{align}
\Psi(x-x_p)=\delta(x-x_p),
\end{align}
so that we may easily invert for the expansion coefficient $\hat{a}^i_j$ in the functional derivative and obtain
\begin{align}\label{eq:discrete_functional_derivative}
\frac{\delta \fa}{\delta f}\bigg|_{f_h}=\sum_p\sum_{i\in I}\sum_{j\in J^i}\frac{\partial A}{\partial w_p}\MV^{i\dagger}_{jp}\mathbf{1}_{\Omega_i}(x_p)\mathbf{1}_{\Omega_i}(x)\mathsf{V}^i_j(v).
\end{align}
The set of pseudo inverse operators $\{\MV^{i\dagger}_{jp}\}_{i\in I}$, one for each domain $\Omega_i$, are required to satisfy the conditions
\begin{align}
\sum_{\{p\, |\, x_p\in\Omega_i\}}\MV^i_{jp}\MV^{i\dagger}_{\ell p}=\delta_{j\ell}, \qquad \forall\, \{i,j,\ell \, |\, i\in I,(j,\ell)\in J^i\}.
\end{align}
with the matrices $\MV^i_{pj}$ defined according to
\begin{align}
\MV^i_{jp}=\int \Phi(v-v_p)\mathsf{V}^i_j(v)dv, \qquad \forall\, \{i,j,p\, |\, i\in I,j\in J^i,x_p\in\Omega_i\}.
\end{align}
The pseudo-inverses can be constructed, e.g., as the Moore-Penrose matrices, as long as $\MV^i_{jp}$ remain of full rank. This can be achieved with an adaptive meshing of the velocity space mindful of the particle locations $v_p$. For a more detailed analysis on discretizing the functional derivative, we refer the reader to Appendix \ref{app:discretization}.

\section{Finite-dimensional metric bracket and Casimir invariants}
\label{sec:finite-bracket}
Throughout this Section, the indices $(i,k)$ refer to degrees of freedom in the  $x$-space, $(j,\ell)$ refer to the degrees of freedom in the $v$-space, and indices $(p,\bar{p})$ are the macro particle indices.  

To obtain a finite dimensional bracket $(\fa_h,\fb_h)_h=(\fa,\fb)[f_h]$, we simply insert the finite-dimensional expression for the functional derivative into the infinite-dimensional bracket 
\begin{align}
(\fa_h,\fb_h)_h(w)=\sum_{p,\bar{p}}\sum_{(i,k)\in I}\sum_{j\in J^i}\sum_{\ell\in J^k}\mathbf{1}_{\Omega_i}(x_p)\mathbf{1}_{\Omega_k}(x_{\bar{p}})\frac{\partial A}{\partial w_p}\MV^{i\dagger}_{jp}\ML^{ik}_{j\ell}(w)\MV^{k\dagger}_{\ell \bar{p}}\frac{\partial B}{\partial w_{\bar{p}}}.
\end{align}
The $\ML$ in the core of the bracket is defined according to
\begin{multline}
\ML^{ik}_{j\ell}(w)=\int\int\left(\mathbf{1}_{\Omega_i}(x)\frac{\partial\mathsf{V}^i_j(v)}{\partial v}-\mathbf{1}_{\Omega_i}(x')\frac{\partial\mathsf{V}^i_j(v')}{\partial v'}\right)\\
\cdot W_h(w;z,z')\cdot\left(\mathbf{1}_{\Omega_k}(x)\frac{\partial \mathsf{V}^k_{\ell}(v)}{\partial v}-\mathbf{1}_{\Omega_k}(x')\frac{\partial \mathsf{V}^k_{\ell}(v')}{\partial v'}\right)dzdz'.
\end{multline}
In defining the numerical correspondent $W_h$ for the dyad $W$, it would be rather inconvenient to use the expression $M(f_h)=f_h$ with the choice for the shape function $\Psi(x-x_p)=\delta(x-x_p)$. Due to the $\delta(x-x')$ in $W$, the finite-dimensional bracket would vanish identically unless the two colliding particles were exactly at the same location in $x$. Thus we choose $M$ to be a projection, 
\begin{align}
M(f_h)(z)=\sum_p\sum_{i\in I}\frac{w_p\mathbf{1}_{\Omega_i}(x_p)}{V(\Omega_i)}\mathbf{1}_{\Omega_i}(x)\Phi(v-v_p),
\end{align}
with $V(\Omega_i)=\int \mathbf{1}_{\Omega_i}(x)dx$ so that $\sum_p w_p\mathbf{1}_{\Omega_i}(x_p)/V(\Omega_i)$ corresponds to the density within the $x$-space domain $\Omega_i$. 

Now the function $\delta(x-x')$ in the dyad $W$ becomes useful. We find that the tensor $\ML$ becomes diagonal with respect to the $x$-space domains $\Omega_i$
\begin{align}
\ML^{ik}_{j\ell}(w)=\delta_{ik}\ML^i_{j\ell}(w),
\end{align}
where for each domain $\Omega_i$, we have $\ML^i_{j\ell}(w)$ given by
\begin{align}
\ML^i_{j\ell}(w)=\int\int\left(\frac{\partial \mathsf{V}^i_j(v)}{\partial v}-\frac{\partial \mathsf{V}^i_j(v')}{\partial v'}\right)\cdot W_h^i(w;v,v')\cdot \left(\frac{\partial \mathsf{V}^i_j(v)}{\partial v}-\frac{\partial \mathsf{V}^i_j(v')}{\partial v'}\right)dvdv',
\end{align}
and the dyad $W_h^i$ defined according to
\begin{align}
W_h^i(w;v,v')=-Q(v-v')\frac{V(\Omega_i)}{2}
\sum_{p,\bar{p}}\frac{w_pw_{\bar{p}}\mathbf{1}_{\Omega_i}(x_p,x_{\bar{p}})}{V(\Omega_i)V(\Omega_i)}
\Phi(v-v_p)\Phi(v'-v_{\bar{p}}),
\end{align}
where $\mathbf{1}_{\Omega_i}(x_p,x_{\bar{p}})=\mathbf{1}_{\Omega_i}(x_p)\mathbf{1}_{\Omega_i}(x_{\bar{p}})$ is a short-hand notation. This simplifies the finite-dimensional bracket into a block diagonal form with respect to the particle positions $x_p$ and $x_{\bar{p}}$ in the sense that they both have to reside within the same domain $\Omega_i$,
\begin{align}\label{eq:finite-bracket}
(\fa_h,\fb_h)_h(w)=\sum_{i\in I}\sum_{p,\bar{p}}\sum_{(j,\ell)\in J^i}\mathbf{1}_{\Omega_i}(x_p,x_{\bar{p}})\frac{\partial A}{\partial w_p}\MV^{i\dagger}_{jp}
\ML^{i}_{j\ell}
\MV^{i\dagger}_{\ell \bar{p}}\frac{\partial B}{\partial w_{\bar{p}}}.
\end{align}

The other significant benefit of our choice is the preservation of the finite-dimensional versions of the Casimirs. To prove our word, let us consider the quantity
\begin{align}
\fc=\int g(x)\phi(v)f(z)dz,
\end{align}
which, with choices $\phi(v)=(1,v,|v|^2/2)$ would correspond to the Casimirs $(\fc_\fm,\fc_\fp,\fc_\fe)$ of the infinite-dimensional system. In terms of our chosen basis, $\fc$ has the finite-dimensional approximation 
\begin{align}
\fc_h&=\int \sum_{i\in I}\sum_{j\in J^i}\hat{g}^i\mathbf{1}_{\Omega_i}(x)\hat{\phi}^i_j\mathsf{V}^i_j(v)f_h(z)dz=\sum_p\sum_{i\in I}\sum_{j\in J^i}w_p \hat{g}^i\mathbf{1}_{\Omega_i}(x_p)\hat{\phi}^i_j\mathbb{V}^i_{jp}.
\end{align}
Inserting this expression and an arbitrary finite-dimensional functional $\fa_h$ into the finite-dimensional bracket, we obtain
\begin{align}
(\fc_h,\fa_h)_h(w)=\sum_{i\in I}\sum_{\bar{p}}\sum_{(j,\ell)\in J^i}\hat{g}^i\hat{\phi}^i_j
\ML^{i}_{j\ell}(w)
\mathbf{1}_{\Omega_i}(x_{\bar{p}})\MV^{i\dagger}_{\ell\bar{p}}\frac{\partial A}{\partial w_{\bar{p}}},
\end{align}
where of specific interest to us are the quantities
\begin{multline}
\sum_{j\in J^i}\hat{\phi}^i_j\ML^i_{j\ell}(w)=\int\int\left(\frac{\partial \sum_j\hat{\phi}^i_j\mathsf{V}^i_j(v)}{\partial v}-\frac{\partial \sum_j\hat{\phi}^i_j\mathsf{V}^i_j(v')}{\partial v'}\right)\\
\cdot W_h^i(w;v,v')\cdot \left(\frac{\partial \mathsf{V}^i_{\ell}(v)}{\partial v}-\frac{\partial \mathsf{V}^i_{\ell}(v')}{\partial v'}\right)dvdv', \quad \forall\, \{i,\ell|i\in I,\ell\in J^i\}.
\end{multline}
Since the bases $\{\mathsf{V}^i_j\}_{j\in J^i}$ in each domain $\Omega_i$ were chosen so that polynomials up to second order can be represented exactly, we find 
\begin{align}
\sum_{j\in J^i}\hat{1}^i_j\ML^i_{j\ell}(w)&=\int\int\left(\frac{\partial 1}{\partial v}-\frac{\partial 1}{\partial v'}\right)\cdot W_h^i(w;v,v')\cdot \left(\frac{\partial \mathsf{V}^i_{\ell}}{\partial v}-\frac{\partial \mathsf{V}^i_{\ell}}{\partial v'}\right)dvdv' = 0,
\\
\sum_{j\in J^i}\hat{v}^i_j\ML^i_{j\ell}(w)&=\int\int\left(\frac{\partial v}{\partial v}-\frac{\partial v'}{\partial v'}\right)\cdot W_h^i(w;v,v')\cdot \left(\frac{\partial \mathsf{V}^i_{\ell}}{\partial v}-\frac{\partial \mathsf{V}^i_{\ell}}{\partial v'}\right)dvdv'=0,
\\
\sum_{j\in J^i}\hat{e}^i_j\ML^i_{j\ell}(w)&=\int\int\frac{1}{2}\left(\frac{\partial |v|^2}{\partial v}-\frac{\partial |v'|^2}{\partial v'}\right)\cdot W_h^i(w;v,v')\cdot \left(\frac{\partial \mathsf{V}^i_{\ell}}{\partial v}-\frac{\partial \mathsf{V}^i_{\ell}}{\partial v'}\right)dvdv'=0.
\end{align}
Thus, for functions $\phi(v)=(1,v,|v|^2/2)$, for which the degrees of freedom are given exactly by $\hat{\phi}^i_j=(\hat{1}^i_j,\hat{v}^i_j,\hat{e}^i_j)$, we have $(\fc_h,\fa_h)_h(w)=0$ for arbitrary values of $\hat{g}^i$. This confirms that our finite-dimensional metric bracket has Casimirs corresponding to mass density, kinetic momentum density, and kinetic energy density, in a similar manner as in the finite-dimensional system.

In order to verify that the finite-dimensional bracket \eqref{eq:finite-bracket} specifies a finite-dimensional metriplectic system, with conservation of the Hamiltonian, dissipation of free energy, and production of entropy, the finite-dimensional version of the Hamiltonian must be a Casimir of the bracket and the bracket must be negative semi-definite. Since the $f$-dependent part of the Vlasov-Maxwell Hamiltonian is reproduced with the choices $\hat{g}^i=1$, the discrete Hamiltonian indeed is a Casimir of the finite-dimensional metric bracket. The dissipation of free energy, on the other hand, is guaranteed as the finite-dimensional metric bracket is negative semi-definite by definition. Conversely, the rate of change of entropy will remain non-negative. 

\section{Finite-dimensional expression for entropy}
\label{sec:entropy}
While the properties of the finite-dimensional metric bracket and the existence of the finite-dimensional Casimirs are guaranteed regardless of the specific form chosen for the discrete entropy functional, the expression should nevertheless be chosen to closely approximate the true entropy functional and to simultaneously remain a Casimir of the finite-dimensional Vlasov-Maxwell Poisson bracket. This way we ensure the compatibility of the metric bracket with the existing particle-in-cell Poisson integrators. 

We propose the following expression for the finite-dimensional entropy functional
\begin{align}\label{eq:finite-dimensional-entropy}
S(w)=-\sum_p w_p \text{ln}\,w_p,
\end{align}
which is trivially a Casimir of the existing Poisson brackets as they operate on $(x_p,v_p)$ and not on $w_p$. To justify our proposed expression for $S(w)$, we first estimate the true entropy functional with respect to a generic macro particle representation for the distribution function. Using a short-hand notation $\gamma_p(x,v)=\Psi(x-x_p)\Phi(v-v_p)$ leads to
\begin{align}
S[f_h]&=-\sum_p\int w_p \gamma_p\text{ln}\left(\sum_{\bar{p}} w_{\bar{p}}\gamma_{\bar{p}}\right)dz\nonumber\\
&\approx -\sum_p\int w_p \gamma_p \text{ln}(w_p\gamma_p)dz\nonumber\\
&=-\sum_p w_p\text{ln} w_p-\sum_p w_p \int \gamma_p\text{ln}\gamma_p\,dz\nonumber\\
&=-\sum_p w_p \text{ln}\,w_p-\sum_p w_p \int \Psi(x)\Phi(v)\text{ln}(\Psi(x)\Phi(v))\,dz
\end{align}
where the approximation in the second line is based on the assumption that within the support of $\gamma_p(x,v)=\Psi(x-x_p)\Phi(v-v_p)$, the value of $\gamma_{\bar{p}}(x,v)$ is close to zero for all $\bar{p}\neq p$. In other words, we assume that the macro particles don't overlap much. On the last line we have used the translation invariance of the phase space volume element $dz=dxdv$. 

The assumption of non-overlapping macro particles is very accurate especially if delta-functions are used for the shape functions. However, the use of delta functions in the above expression of entropy is not entirely rigorous: the integral on the last line when evaluated with respect to Gaussian shape functions and letting the width of the Gaussians to approach zero, approximating a delta function, approaches infinity. Fortunately, the last, problematic term in the entropy is a Casimir of both the existing Poisson brackets and the metric bracket proposed in the previous section. Thus, regardless of the value of the integral or the weights, the last term does not contribute to the dynamics. This leads us to conjecture that the expression \eqref{eq:finite-dimensional-entropy} is in fact the correct entropy for generic macro particle discretizations. It is also physically intuitive, requiring strict non-negativity of macro particle weights.

\section{Temporal discretization with Discrete Gradient methods}
So far, we have managed to convert the infinite-dimensional metric bracket representative of the Landau collision operator into a finite-dimensional metric bracket using a marker representation for the distribution function. We have also demonstrated the existence of finite-dimensional Casimirs corresponding to the mass, kinetic momentum, and kinetic energy density, and the dissipation of the free energy and production of entropy. Next we seek for temporal discretizations that preserve these properties. The specific property that the finite-dimensional Casimirs of the metric bracket are linear with respect to the marker weights turns out to be useful for success.

In terms of the finite-dimensional metric bracket, the time-continuous equations for the marker weights are obtained from
\begin{align}
\frac{dw_p}{dt}=(w_p,\ff_h)_h=\sum_{\bar{p}}\MG_{p\bar{p}}(w)\frac{\partial\ff_h(w)}{\partial w_{\bar{p}}},
\end{align}
where the matrix $\MG(w)$ is given by
\begin{align}
\MG_{p\bar{p}}(w) = \sum_{i\in I}\sum_{(j,\ell)\in J^i}\mathbf{1}_{\Omega_i}(x_p,x_{\bar{p}})\MV^{i\dagger}_{jp}\ML^{i}_{j\ell}(w)
\MV^{i\dagger}_{\ell\bar{p}}.
\end{align}
We discretize this system in time exploiting the so-called discrete gradient methods \cite{Quispel:1996,McLachlan:1999,QuispelMcLaren:2008,MansfieldQuispel:2009,CohenHairer:2011}. Using the notation $w=(\dots,w_p,\dots)$ and $\nabla\fa_h=(\dots,\partial\fa_h/\partial w_p,\dots)$, the temporal discretization is expressed compactly as
\begin{align}
\frac{w^1-w^0}{\Delta t}=\MG(w^{1/2})\cdot\overline{\nabla\ff}_h(w^1,w^0),
\end{align}
where $w^{1/2}=(w^1+w^0)/2$ and the operator $\overline{\nabla \fa}_h$ is the discrete gradient of the function $\fa_h$ required to satisfy the conditions
\begin{align}
(w^1-w^0)\cdot\overline{\nabla \fa}_h(w^1,w^0)&=\fa_h(w^1)-\fa_h(w^0),\\
\overline{\nabla \fa}_h(w,w)&=\nabla\fa_h(w).
\end{align}
Several such construction are known in the literature.

To demonstrate that the chosen temporal discretization method preserves the Casimirs and guarantees the dissipation of free energy as well as the production of entropy, we first use the defining property of the discrete gradient and the chosen temporal stepping method with respect to an arbitrary functional
\begin{align}
\fa_h(w^1)-\fa_h(w^0)&=\overline{\nabla\fa}_h(w^1,w^0)\cdot(w^1-w^0)\nonumber\\
&=\Delta t\, \overline{\nabla\fa}_h(w^1,w^0)\cdot\MG(w^{1/2})\cdot\overline{\nabla\ff}_h(w^1,w^0).
\end{align}
The finite-dimensional Casimirs $\fc_h$ of the bracket $(\fa_h,\fb_h)_h$ are linear with respect to the marker weights. Secondly, they satisfy the condition $\nabla\fc_h\cdot\MG(w)=0$ where $\nabla\fc_h$ is a constant vector independent of $w$. As the discrete gradient is exact for linear functions, we immediately obtain
\begin{align}
\fc_h(w^1)-\fc_h(w^0)=\Delta t\,\nabla\fc_h\cdot\MG(w^{1/2})\cdot\overline{\nabla\ff}_h(w^1,w^0)=0.
\end{align}
In a matter of fact, this quite useful result can be reproduced also for quadratic Casimirs~\cite{CohenHairer:2011}.
As the $f$-dependent part of the Vlasov-Maxwell Hamiltonian can be expressed in terms of these Casimirs, the method guarantees also the presrvation of the Hamiltonian. The dissipation of free energy follows from the assumed negative semi-definiteness of the metric bracket
\begin{align}
\ff_h(w^1)-\ff_h(w^0)=\Delta t\, \overline{\nabla\ff}_h(w^1,w^0)\cdot\MG(w^{1/2})\cdot\overline{\nabla\ff}_h(w^1,w^0)\leq 0,
\end{align}
while the non-negative rate of entropy production follows from the invariance of the Hamiltonian and the dissipation of free energy
\begin{align}
\fs_h(w^1)-\fs_h(w^0)=\ff_h(w^0)-\fh_h(w^0)-\ff_h(w^1)+\fh_h(w^1)=\ff_h(w^0)-\ff_h(w^1)\geq 0.
\end{align}
\section{Summary and discussion}
In this paper, we have described a new framework for addressing the nonlinear Landau collision operator in terms of particle-in-cell methods. Based on the underlying metriplectic structure of the collision operator, we used a macro particle discretization for the distribution function and transformed the infinite-dimensional metriplectic formulation of the Landau collision operator into a finite-dimensional time-continuous metriplectic system for advancing the macro particle weights. The finite-dimensional bracket was then shown to have Casimir invariants corresponding to mass, kinetic momentum, and kinetic energy density, analogously to the infinite-dimensional system. Finally, temporal discretization was accomplished using the concept of discrete gradients, preserving the Casimirs and providing dissipation of free energy as well as production of entropy. 

These results were largely possible due to two essential ingredients: (1) the pseudo-inverse definition of the discrete functional derivative with respect to a macro particle distribution function relative to a phase space Galerkin basis, and (2) making the wise choice of demanding that the phase space Galerkin basis contains the functions $(1,v,|v|^2)$. We would also like to note that the pseudo-inverse definition of the functional derivative works even if we do not demand that the associated Galerkin basis resolves $(1,v,|v|^2)$. In certain cases this might be useful in the sense that the metric bracket would produce entropy, in spite of losing energy and momentum conservation.

A possible limitation of the proposed algorithm is that it preserves Hamiltonian trajectories. That is to say that there is no scattering in velocity space of the macro particle locations, implying that phase space density cannot be transferred to areas of phase space where there are no macro particles. 
This issue can be mitigated by populating a sufficiently large area of velocity space with macro particles and by adjusting the velocity space meshing. For example, the initial particle sampling could be uniform in velocity space up to some sufficiently large maximum velocity, with weights set so to satisfy the actual initial conditions for the distribution function.

Although the presented results open up a new and exciting research direction, they provide only the first step. For example, one of the remaining open questions is to find a metriplectic algorithm that can be algebraically shown to retain the strict non-negativity of the macro particle weights, a condition that is necessary for a meaningful definition of entropy. We leave this and other such challenges to future work.

\begin{acknowledgments}
EH was supported by the U.S. Department of Energy Contract No. DE-AC02-09-CH11466. MK received funding from the European Union's Horizon 2020 research and innovation programme under the Marie Sklodowska-Curie grant agreement No 708124. The research of JWB was funded via the U.S. Department of Energy grant No. DE-AC05-06OR23100 and carried under the Fusion Energy Sciences Postdoctoral Research Program, administered by the Oak Ridge Institute for Science and Education (ORISE), managed by Oak Ridge Associated Universities (ORAU) for DOE. The views and opinions expressed herein do not necessarily reflect those of the European Commission or the U.S. Department of Energy.
\end{acknowledgments}



\appendix
\section{Macro-particle discretization of functional derivatives}
\label{app:discretization}
In discretizing the functional derivatives, we require the following equivalence between the continuous and discrete functional derivatives,
\begin{align}
\bracket{ \dfrac{\delta \fa}{\delta f} [f_h] , \, g_h }_{L^{2}}
=
\bracket{ \dfrac{\partial \fa_h}{\partial w} , \, \bar{w} }_{\mathbb{R}^{N}} ,
\end{align}
with $\bracket{\cdot,\cdot}_{L^2}$ an $L^2$ inner product, $\bracket{\cdot,\cdot}_{\mathbb{R}^N}$ an ordinary vector dot product, and the test function $g_h$ given by
\begin{align}
g_h (t,x,v) = \sum \limits_p \, \Psi(x-x_p) \, \Phi(v-v_p) \, \bar{w}_p (t) .
\end{align}
This expands into
\begin{align}
\sum \limits_p \int \dfrac{\delta \fa}{\delta f}\bigg\vert_{f_h} \, \Psi(x-x_p) \, \Phi(v-v_p) \, \bar{w}_p \, dx \, dv
=
\sum \limits_p \dfrac{\partial \fa_h}{\partial w_p} \, \bar{w}_p .
\end{align}
Next we express the functional derivative of $\fa$ in the basis
\begin{align}
\dfrac{\delta \fa}{\delta f}\bigg|_{f_h} = \sum \limits_{i\in I}\sum \limits_{j\in J^i} \hat{a}_{j}^{i} \, \mathbf{1}_{\Omega_i}(x) \, \mathsf{V}^i_j(v) ,
\end{align}
so that
\begin{align}
\sum \limits_p  \sum \limits_{i\in I}\sum \limits_{j\in J^i} \int \hat{a}_{j}^{i} \, \mathbf{1}_{\Omega_i}(x) \, \mathsf{V}^i_j(v) \, \Psi(x-x_p) \, \Phi(v-v_p) \, \bar{w}_p \, dx \, dv
=
\sum \limits_p \dfrac{\partial \fa_h}{\partial w_p} \, \bar{w}_p .
\end{align}
Then defining the matrices
\begin{align}
\MV^i_{jp} = \int \limits_{\Omega_i} \Phi(v-v_p) \, \mathsf{V}^i_j(v) \, dv , \qquad \forall \, \{ i,j,p \, |\, i\in I, j\in J^i, x_p \in \Omega_i \} ,
\end{align}
we obtain 
\begin{align}
\sum \limits_p \sum \limits_{i\in I}\sum \limits_{j\in J^i} \int \hat{a}_{j}^{i} \, \mathbf{1}_{\Omega_i}(x) \, \MV^i_{jp} \, \Psi(x-x_p) \, \bar{w}_p \, dx
=
\sum \limits_p \dfrac{\partial \fa_h}{\partial w_p} \, \bar{w}_p .
\end{align}
Choosing the $x$-space shape function to be $\Psi(x-x_p) = \delta(x-x_p)$, we get
\begin{align}
\sum \limits_p \sum \limits_{i\in I}\sum \limits_{j\in J^i} \hat{a}_{j}^{i} \, \mathbf{1}_{\Omega_i}(x_p) \, \MV^i_{jp} \, \bar{w}_p
=
\sum \limits_p \dfrac{\partial \fa_h}{\partial w_p} \, \bar{w}_p .
\end{align}
As the coefficients $\bar{w}_p$ of the test function $g_h$ are arbitrary, this is equivalent to
\begin{align}
\sum \limits_{i\in I}\sum \limits_{j\in J^i} \hat{a}_{j}^{i} \, \mathbf{1}_{\Omega_i}(x_p) \, \MV^i_{jp}
=
\dfrac{\partial \fa_h}{\partial w_p}
, \qquad \forall \, p .
\end{align}
Next pick $i\in I$ such that $x_{p} \in \Omega_i$ and note that for $x_{p} \notin \Omega_i$ we have $\MV^i_{jp} = 0 \; \forall j\in J^i$, hence we can write
\begin{align}
\sum \limits_{j\in J^i} \hat{a}_{j}^{i}  \, \MV^i_{jp}
=
\dfrac{\partial \fa_h}{\partial w_p}
, \qquad \forall \, \{ i,p \, |\, i\in I, x_p \in \Omega_i \}  .
\end{align}
Introducing the pseudo inverse operators $\{\MV^{i\dagger}_{jp}\}_{i\in I}$, one for each domain $\Omega_i$, requiring
\begin{align}
\sum_{\{p\, |\, x_p\in\Omega_i\}}\MV^i_{jp}\MV^{i\dagger}_{\ell p}=\delta_{j\ell}, \qquad \forall\, \{i,j,\ell,p\, |\, i\in I,(j,\ell)\in J^i,x_p\in\Omega_i\},
\end{align}
multiplying the previous expression with $\MV^{i\dagger}_{\ell p}$, and summing over $p$, we get for each $i$
\begin{align}
\sum \limits_{j\in J^i} \sum_{\{p\, |\, x_p\in\Omega_i\}} \hat{a}_{j}^{i} \, \MV^i_{jp}\MV^{i\dagger}_{\ell p}
= \hat{a}_{\ell}^{i}
= \sum_{\{p\, |\, x_p\in\Omega_i\}} \dfrac{\partial \fa_h}{\partial w_p} \MV^{i\dagger}_{\ell p} \quad \forall \{i,\ell \, |\, i\in I, \ell\in J^i\}.
\end{align}
With that, the discrete functional derivative becomes
\begin{align}
\dfrac{\delta \fa}{\delta f}\bigg|_{f_h}
&= \sum \limits_{i\in I}\sum \limits_{j\in J^i} \sum_{\{p\, |\, x_p\in\Omega_i\}} \dfrac{\partial \fa_h}{\partial w_p} \, \MV^{i\dagger}_{jp} \, \mathbf{1}_{\Omega_i}(x) \, \mathsf{V}^i_j(v)\nonumber\\
&= \sum \limits_{i\in I}\sum \limits_{j\in J^i} \sum_{p} \dfrac{\partial \fa_h}{\partial w_p} \, \MV^{i\dagger}_{jp} \, \mathbf{1}_{\Omega_i}(x_p) \, \mathbf{1}_{\Omega_i}(x) \, \mathsf{V}^i_j(v) ,
\end{align}
which is the expression introduced in the main body of the paper.

\bibliographystyle{apsrev4-1}
\bibliography{bibfile}   
\end{document}